\documentclass[epj]{svjour}
\usepackage{graphicx}
\title{A Cluster Monte Carlo Algorithm for 2-Dimensional Spin Glasses}
\author{J. Houdayer}
\institute{Institut f\"{u}r Physik, D-55099 Mainz, Germany,
\email{jerome.houdayer@polytechnique.org}}
\begin{document}
\abstract{A new Monte Carlo algorithm for 2-dimensional spin glasses
is presented. The use of clusters makes possible global updates and
leads to a gain in speed of several orders of magnitude. As an
example, we study the 2-dimensional $\pm J$ Edwards-Anderson
model. The new algorithm allows us to equilibrate systems of size
$100^2$ down to temperature $T=0.1$. Our main result is that the
correlation length diverges as an exponential ($\xi\sim e^{2\beta J}$)
and not as a power law as $T \to T_c = 0$.
}
\maketitle

\section{Introduction}
\subsection{General considerations}
The understanding of (disordered) Ising ferromagnets has been greatly
enhanced by fast Monte Carlo (MC) simulations using cluster algorithms
\cite{SwendsenWang87,Wolff89}. Unfortunately those techniques cannot be
directly applied to models such as spin glasses (SG)
because of frustration. Attempts have been made to generalise cluster
methods \cite{Liang92,CataudellaFranzese94,MatsubaraSato97} but the
resulting algorithms are complicated and the speed increase is not
impressive. A cluster algorithm for fully frustrated systems already
exists~\cite{KandelBenav90} but cannot tackle disorder. Other
techniques such as exchange MC (EMC) (also called parallel tempering)
\cite{HukushimaNemoto96} allow big improvements over standard
one-spin flip MC and are widely used for SG. Nevertheless the sizes
and temperatures accessible to simulations are still not enough to
clearly solve many important issues (see~\cite{MarinariParisi98c} for
a review). In the case of 2-dimensional spin glasses, the best method
to date is the replica MC (RMC) by Swendsen and
Wang~\cite{SwendsenWang86} (which essentially reduces to EMC in higher
dimensions).

We present here a new cluster MC algorithm for 2-dimensional SG which
is much faster than previous MC techniques (namely RMC). It thus gives
access to sizes and temperatures which were unreachable before. Note
that transfer matrix methods which are widely used for 2-dimensional
systems are limited to ``small'' sizes, usually no more than
$16\times\infty$ which appears to be not enough to tackle certain open
problems. With our new tool we have studied the SG transition in the
2-dimensional $\pm J$ Edwards-Anderson (EA)~\cite{EdwardsAnderson75}
model for which several questions are still unsettled. In particular
the value of the critical temperature (zero or not)
\cite{BhattYoung88,LemkeCampbell96,ShirakuraMatsubara96,KitataniSinada00,ShirakuraMatsubara00} and the nature of the
divergences (power laws or exponentials)~\cite{SaulKardar93} are still
debated. We present evidence at the end of this article that $T_c=0$
and that the correlation length follows an exponential law $\xi \sim
e^{2\beta J}$, which is different from the standard lore.

\subsection{Models}
The model we consider here is a general Ising spin model on a lattice,
whose Hamiltonian is:
\begin{equation}
H(S) = - \sum_{i,j} J_{ij} S_i S_j - \sum_i h_i S_i,
\label{eq_hamilton}
\end{equation}
where $S_i=\pm 1$. The interactions $J_{ij}$ and the magnetic
fields $h_i$ are {\it any} fixed real numbers. Let $N$ be the number
of spins. Although the algorithm described below is correct in all
cases, it is faster than more traditional methods only in the
2-dimensional case with nearest neighbour interactions. This allow
to study the Ising spin glass, the (disordered) Ising ferromagnet and
the random field Ising model in two dimensions. We will only consider
those cases in the following.
\section{The algorithm}
\subsection{The cluster Monte Carlo step}
Let us now describe a Monte Carlo move which allows a global update of
the spin configurations. Consider a system consisting of two
independent spin configurations at the same temperature $1/\beta$;
thus a ``configuration'' of this system is a set of two spin
configurations: ${\cal C} = (\{S_i^1\},\{S_i^2\})$. The same
Hamiltonian (the same $J_{ij}$'s and $h_i$'s) apply to both
configurations. We want to sample the configurations with the weight:
\begin{equation}
P({\cal C}) \propto \exp \left[-\beta \left(H(S^1)+H(S^2)\right)\right].
\end{equation}
To do this, it is sufficient to have an ergodic Markov chain and to
enforce the ``detailed balance'' condition:
\begin{equation}
P({\cal C})\Pi_{\cal C\rightarrow C'}=P({\cal C'})\Pi_{\cal C'\rightarrow
C},
\label{eq_det_bal}
\end{equation}
where $\Pi_{\cal C\rightarrow C'}$ is the transition probability from
$\cal C$ to $\cal C'$.

\begin{figure}
\begin{center}
\resizebox{0.9\linewidth}{!}{\includegraphics{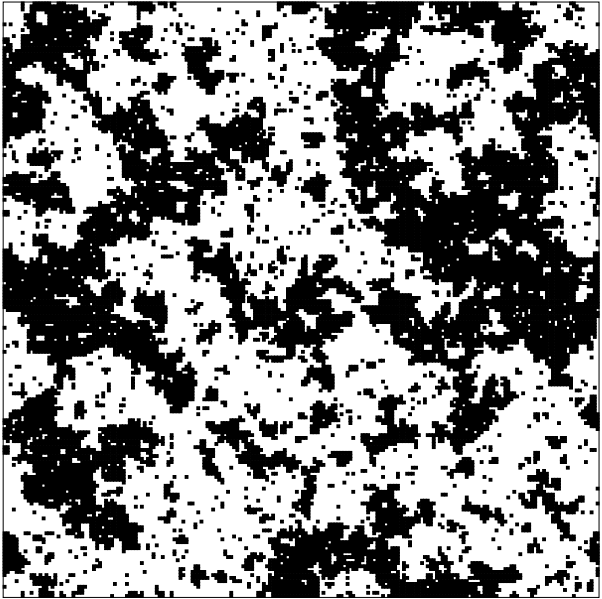}}
\end{center}
\caption{An example of the value of the $q_i$'s ($1$ is white, $-1$ 
is black). The system is here a $200^2$ $\pm 1$ 2-dimensional
Edwards-Anderson spin glass at equilibrium at $\beta=1.75$.}
\label{fig_qi}
\end{figure}

We define the local overlap at site $i$ between the two replicas by
$q_i=S_i^1S_i^2$. This defines two domains on the lattice, the sites
with $q_i=1$ and the sites with $q_i=-1$. We call {\it clusters} the
connected parts of these domains (two sites $i$ and $j$ are connected
if $J_{ij}\neq 0$). Our cluster Monte Carlo step proceeds as follow:
choose one site at random for which $q_i=-1$, and flip the cluster to
which it belongs in {\it both} configurations. It is quite easy to
check that $H(S^1)+H(S^2)$ is unchanged by this transformation, the
interface and magnetic energies of the cluster in both configurations
being exchanged by the flip. Moreover the $q_i$'s are also unchanged,
together with the definition of the clusters. It is thus clear that
equation~\ref{eq_det_bal} is verified for this step since $P({\cal
C})=P({\cal C'})$ and $\Pi_{\cal C\rightarrow C'}=\Pi_{\cal
C'\rightarrow C}$. Note that in the case where the $h_i$'s are zero,
there is no magnetic energy and one can also choose a spin with
$q_i=+1$ to flip a cluster. In figure~\ref{fig_qi} an example of the
values of the $q_i$'s is presented. The clusters are the connected
parts of the white and black domains.

Our construction is very similar to RMC~\cite{SwendsenWang86} with the
essential difference that both replicas have the same
temperature. This point is very important for two reasons: (i) our
cluster moves are accepted with probability one, (ii) we can use more
than two configurations as explained in the next section. The
definition of the clusters is also reminiscent
of~\cite{RednerMachta98} where a cluster \`a la Wolff is grown inside a
region where two replicas are different. This last method works in
random field but requires non frustrated interactions (as far as we
know it has never been extended to frustrated interactions).

\subsection{Description of the algorithm}
In spin systems one is essentially interested in three observables:
the energy of the system ($H$), the magnetisation ($m=\sum S_i/N$) and
for spin glasses the overlap ($q=\sum S_i^1S_i^2/N$). But from the
previous discussion it is clear that the proposed cluster flip keeps
all those quantities constant for the whole system though the energies
and magnetisations of each replica do change. So as described in the
previous section the algorithm is non ergodic. To build a valid Monte
Carlo algorithm, we need first to enforce ergodicity. The simplest way
to achieve this is to add another kind of move: We choose the
standard one-spin flip move with Metropolis acceptance (as done in
RMC). The main point of this new algorithm is that we do not restrict
ourselves to only two configurations. We use $n \gg 2$ replicas at the
same temperature. This trick allows a much faster relaxation because
these $n$ replicas are mixed together very quickly. In particular with
3 replicas, if one flips a cluster between replicas 1 and 2, even if
$q_{12}$ is unchanged, $q_{13}$ and $q_{23}$ do change.

At this point one problem remains, namely the total energy of the $n$
replicas is conserved by the cluster moves. Only the one-spin flip
moves can relax the total energy and the resulting algorithm is much
slower than RMC (and also slower than EMC). That is why we embed our
cluster method in an EMC.

Here are the details of our algorithm: the system is composed of $m$
sets at different temperatures. Each set consists in $n$ independent
replicas at the same temperature. The same Hamiltonian (with the
same $J_{ij}$'s and $h_i$'s) applies to all configurations. The
algorithm repeatedly does the following:
\begin{enumerate}
\item {\bf One-spin flips}: Do a standard one-spin flip move for each
spin in each replica.
\item {\bf Cluster moves}: For each temperature, randomly partition the $n$
replicas in pairs and do one cluster move for each pair.
\item {\bf EMC}: For each pair of neighbouring temperatures, do $n$ standard
EMC updates between the two sets (pairing each replica from one set
with one from the other).
\end{enumerate}
In a standard EMC update, two spin configurations at different
temperatures are exchanged with probability
\begin{equation}
P_{1\leftrightarrow 2} = \min\left(1,e^{(\beta_2-\beta_1)(H_2-H_1)}\right).
\end{equation}

The choice of the $m$ temperatures is made as in EMC: One needs enough
temperatures for the energy distributions of two neighbouring
temperatures to overlap; then the exchange moves can be accepted with
a reasonable probability. A good rule of thumb is to try to fix the
exchange acceptance rate to $1/3$ to optimise the mixing effect. The
choice of $n$ seems to be simple: the larger the better. In fact, the
computation time increases linearly with $n$ but one obtains $n$
independent configurations at each step for each temperature. On the
other hand the mixing effect of the clusters rapidly grows with $n$,
so a large value is preferable, In the following we set $n=32$, the
maximum value allowed by our implementation (spin coding on 4-byte
integers); it would probably be better to use even larger values of $n$.

\begin{figure}
\begin{center}
\resizebox{0.9\linewidth}{!}{\includegraphics{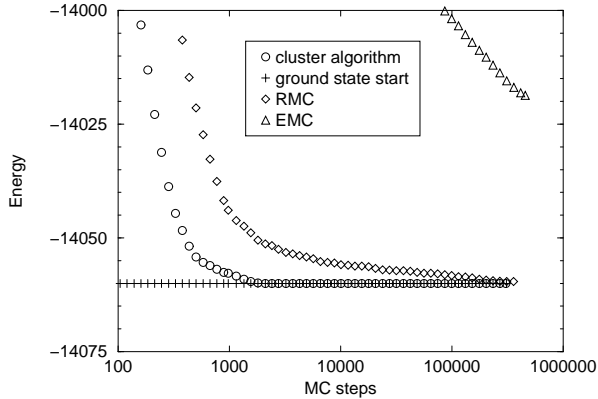}}
\end{center}
\caption{Relaxation of the energy in a $100^2$ spin glass for different
algorithms at $\beta=10$. See the text for details.}
\label{fig_speed_e}
\end{figure}

\begin{figure}
\begin{center}
\resizebox{0.9\linewidth}{!}{\includegraphics{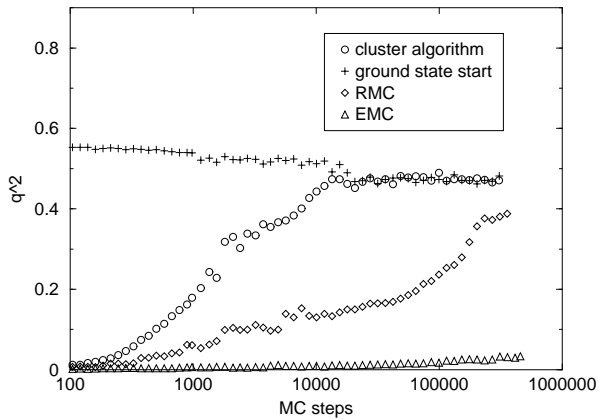}}
\end{center}
\caption{Relaxation of $q^2$ in a $100^2$ spin glass for different algorithms
at $\beta=10$. See the text for details.}
\label{fig_speed_q}
\end{figure}

We investigate the relaxation time on a $100^2$, $\pm 1$ spin glass
($h_i=0$) with periodic boundary conditions at $\beta=10$ using $n=32$
and $m=26$ temperatures in the range $\beta=1\ldots 10$. We
compare our algorithm to EMC and RMC (using the same temperatures)
starting from random configurations and averaging over 32 independent
runs. In figures \ref{fig_speed_e} and
\ref{fig_speed_q}, one can see the dramatic improvement over the EMC
and the big improvement over RMC (at least a factor 100 in this
case). Note that to determine if equilibrium has been reached, we also
reran the algorithm starting from ground state configurations (obtained
from an improved version of the algorithm described
in~\cite{HoudayerMartin99}). The equilibrium value is reached where the
curves meet. It appears that the gain in speed increases with $N$ and
$\beta$ and may become really huge.

Our algorithm is very efficient because the two techniques (clusters
and EMC) are perfectly complementary: somehow the EMC is able to
explore the energy landscape vertically (up and down in energy)
whereas the clusters allow an horizontal search (search at constant
energy). The cluster moves give a very quick mixing of the different
configurations at one temperature and EMC allows for a renewal of the
population. Even if the clusters are the same as in RMC, RMC does not
perform so well because of a less good mixing: In RMC the cluster moves
always happen between the same configurations and they are hampered by
the difference in temperature.

\subsection{Only in 2 dimensions}
Consider what happens in 3 dimensions: During the cluster
construction, usually there are about as many sites with $q_i=1$ as
with $q_i=-1$. In $d=3$ the site percolation threshold is roughly 0.3
and so both $q=1$ and $q=-1$ sites percolate forming two system-size
clusters (and some very small clusters). It is easy to see that
flipping one of those big clusters in both replicas is the same as
exchanging the replicas (except for the very small clusters). Hence,
the move does essentially nothing (for the same reason, RMC becomes
equivalent to EMC in 3 dimensions). This problem is encountered as
soon as the site percolation threshold is less than 0.5 which
essentially forbids all lattices of dimension $d>2$ or with more than
nearest neighbour connections in $d=2$.

\section{The EA model in $d=2$}
Using our algorithm, we have studied the 2-dimensional $J=\pm 1$ EA
model with periodic boundary conditions on square $L\times L$
lattices. We have simulated 4 sizes ($L=12$, 25, 50, 100) for
temperatures ranging from $\beta=0.3$ to $\beta=10$. We have performed
a disorder average over many samples (respectively 400, 400, 200 and
100). The large sizes and low temperatures involved made it difficult
to equilibrate the system so we proceeded as follows. For each sample
we first started from random configurations and waited for the lowest
temperature to be completely occupied by ground states. Then we used
these ground states as initial configurations for all temperatures and
let the algorithm run. The point here is that it is easier for the
algorithm to heat the system than to cool it, since the main obstacle
is the low entropy of low energy states. We respectively used a
relaxation time of $10^3$, $5. 10^3$, $2. 10^4$ and $10^5$ MC steps
for the different sizes.

In each case we gathered statistics for the overlap $q$ and measured the
SG susceptibility
\begin{equation}
\chi=N\overline{\langle q^2\rangle},
\end{equation}
where the over-line denotes the disorder average and the brackets the
thermal average. We also measured the Binder cumulant
\begin{equation}
g = \frac{1}{2}\overline{\left(3-\frac{\langle q^4\rangle}{\langle
q^2\rangle^2}\right)}.
\end{equation}
 
\begin{figure}
\begin{center}
\resizebox{0.9\linewidth}{!}{\includegraphics{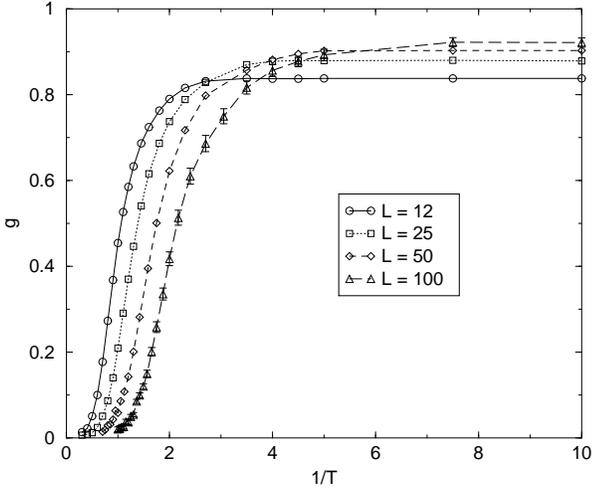}}
\end{center}
\caption{The Binder cumulant $g$ as a function of the inverse
temperature $\beta$. The error bars are shown only for $L=100$, they
are smaller than the symbols for the others sizes.}
\label{fig_g}
\end{figure}

\begin{figure}
\begin{center}
\resizebox{0.9\linewidth}{!}{\includegraphics{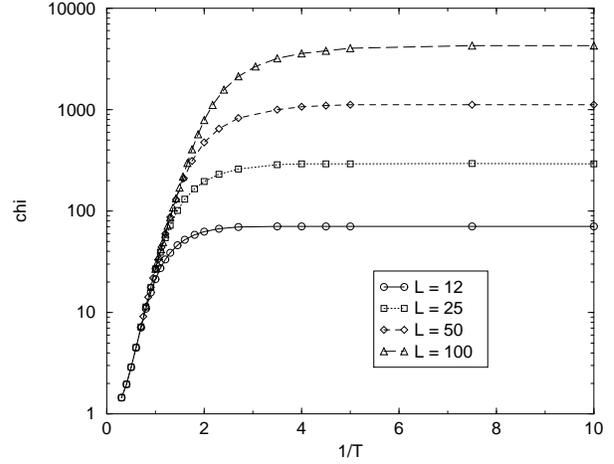}}
\end{center}
\caption{The SG susceptibility $\chi$ as a function of $\beta$. Error
bars are smaller than the symbols.}
\label{fig_chi}
\end{figure}

The standard expected scaling forms~\cite{BhattYoung88} are
$\xi\sim(T-T_c)^{-\nu}$ for the correlation length,
\begin{equation}
\chi\sim L^{2-\eta}\tilde{\chi}(L^{1/\nu}(T-T_c))
\label{eq_pow_chi}
\end{equation}
for the susceptibility and
\begin{equation}
g\sim \tilde{g} (L^{1/\nu}(T-T_c))
\label{eq_pow_g}
\end{equation} 
for the Binder cumulant. Then the value of $g$ should be constant at
the transition and the curves in figure~\ref{fig_g} should cross at
$T=T_c$. The curves definitely cross, but the crossing point goes to
higher and higher values of $\beta$ as the size increases which is a
strong indication that no finite temperature transition occurs
(i.e. $T_c=0$). Even if all the curves have a similar shape, it is
interesting to note that the plateaus at low temperature cannot simply
scale as predicted by equation~\ref{eq_pow_g} (nor by
equation~\ref{eq_exp_g}), since those equations predict that
$g(L,T=0)$ should be a constant (for $T_c=0$) which is clearly not the
case. This behaviour is probably a finite size effect due to the fact
that at these temperatures the system is in its ground state. The
results for $\chi$ are presented in figure~\ref{fig_chi}.

\begin{figure}
\begin{center}
\resizebox{0.9\linewidth}{!}{\includegraphics{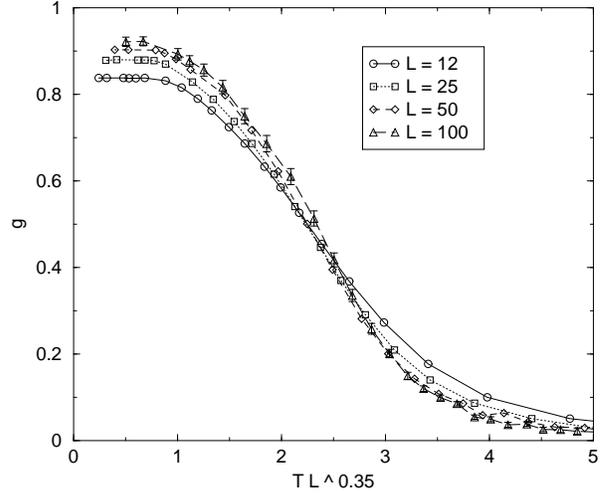}}
\end{center}
\caption{The power law scaling: $g$ as a function of $TL^{0.35}$.}
\label{fig_pow_g}
\end{figure}

\begin{figure}
\begin{center}
\resizebox{0.9\linewidth}{!}{\includegraphics{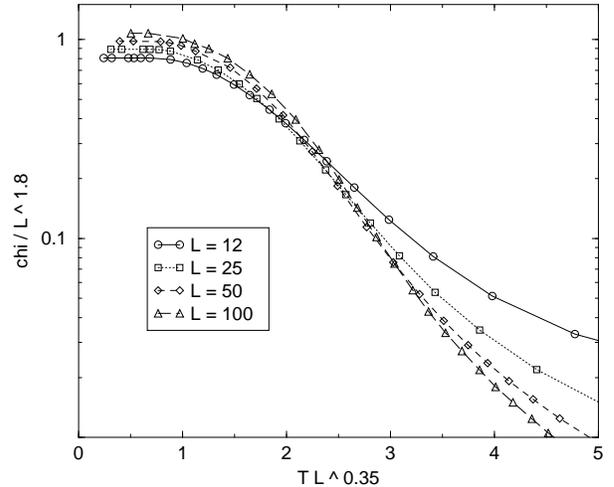}}
\end{center}
\caption{The power law scaling: $\chi/L^{1.8}$ as a function of $TL^{0.35}$.}
\label{fig_pow_chi}
\end{figure}

Figures~\ref{fig_pow_g} and~\ref{fig_pow_chi} show the result of the
scaling of equations~\ref{eq_pow_chi} and~\ref{eq_pow_g} using
$1/\nu=0.35$ and $\eta=0.2$ (and $T_c=0$). In both figures one can see
a clear trend with size and the curves do not to overlap. This effect
is not due to a bad choice of the parameters since the curves on
figure~\ref{fig_pow_g} cross and no horizontal scaling around $0$
could make the curves overlap. The same argument applies to
figure~\ref{fig_pow_chi} for both axes. For the same reason
corrections to scaling as proposed in~\cite{KitataniSinada00} cannot
improve the scaling. This leads us to conclude that this scaling does
not apply.

\begin{figure}
\begin{center}
\resizebox{0.9\linewidth}{!}{\includegraphics{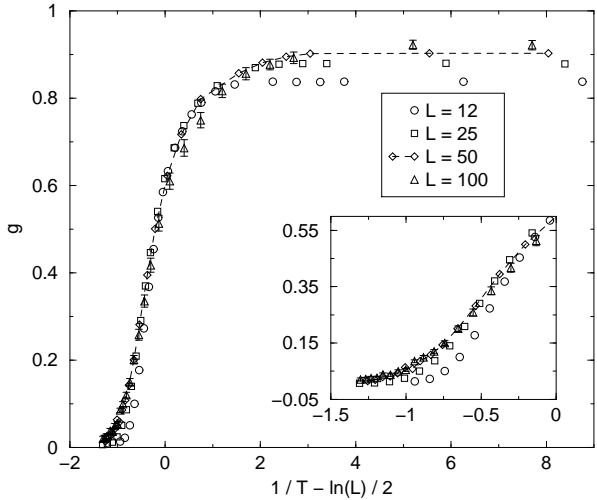}}
\end{center}
\caption{The exponential scaling: $g$ as a function of $\beta-\frac12
\ln L$ (no free parameters). The insert shows an enlargement of the
high temperature region.}
\label{fig_exp_g}
\end{figure}

\begin{figure}
\begin{center}
\resizebox{0.9\linewidth}{!}{\includegraphics{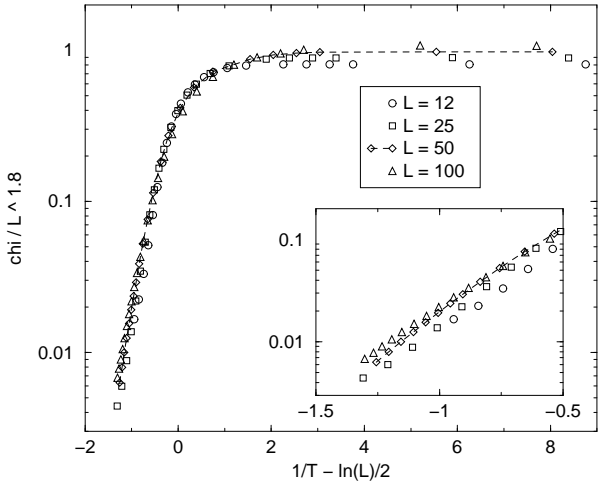}}
\end{center}
\caption{The exponential scaling: $\chi/L^{1.8}$ as a function of
$\beta-\frac12 \ln L$. The insert shows an enlargement of the
high temperature region.}
\label{fig_exp_chi}
\end{figure}

A few years ago, Saul and Kardar~\cite{SaulKardar93} proposed a few
years ago an exponential scaling for the correlation length, namely
$\xi\sim e^{2\beta}$, which leads to:
\begin{equation}
\chi\sim L^{2-\eta}\tilde{\chi}(\beta - \frac12 \ln L),
\label{eq_exp_chi}
\end{equation}
and
\begin{equation}
g\sim \tilde{g} (\beta - \frac12 \ln L).
\label{eq_exp_g}
\end{equation} 
The corresponding scalings are shown in figures~\ref{fig_exp_g}
and~\ref{fig_exp_chi} with $\eta=0.2$. The overlap is much better and
there are fewer free parameters, which seems to indicate that this
scaling is the correct one. In the inserts of these figures on can see
that the discrepancies disappear as the size increases. Finally the
scalings with $T_c\neq 0$ have also been tried (data not shown), but
they give less good results than the exponential scaling though having
two parameters more.

\section{Conclusion}
We have described a cluster Monte Carlo algorithm for Ising spin
models which incorporates an exchange Monte Carlo along with constant
energy ``cluster'' moves. In 2 dimensions with nearest neighbour
interactions, this algorithm allows a gain in speed of several order
of magnitude as compared to replica Monte Carlo and exchange Monte
Carlo. And it works for any kind of interactions and magnetic field
(even frustrated and disordered).

The $\pm J$ 2-dimensional Edwards-Anderson {I}sing spin glass was then
studied at very low temperatures and for large sizes. The critical
temperature appears to be exactly $T_c=0$ and the standard power law
scaling must be replaced by the exponential scaling proposed by Saul
and Kardar~\cite{SaulKardar93} (equations \ref{eq_exp_chi} and
\ref{eq_exp_g}). This probably means that the lower critical dimension is
$d_l=2$ for spin glasses. In an early work~\cite{McMillan84}, McMillan
stated that $d_l>2$ and that at $d=d_l$ one should have $\xi \sim
e^{K\beta ^2}$; we also tried this scaling law but it definitely does
not apply in our case.

As argued in~\cite{HoudayerMartin00b} excitations with non-trivial
topology are necessary to have a spin glass phase, and it is precisely
in this case that the algorithm proposed here does not work. This
point of view is also compatible with~\cite{LemkeCampbell96} where a
2-dimensional spin glass with next-nearest neighbour ferromagnetic
interactions seems to have a finite critical temperature (the
percolation threshold is then less than $1/2$ and excitations with
non-trivial topology are possible).

Only the $\pm J$ spin glass has been studied here; it would be
interesting to see if the spin glass with Gaussian couplings behaves
in the same way; this will be the subject of a publication to come.

\begin{acknowledgement}
The author acknowledges W. Kob, K. Binder and I. Campbell for
fruitful discussions and the Max Planck Institut f\"{u}r
Polymerforschung for its financial support.
\end{acknowledgement}


\begin{thebibliography}{10}
\bibitem{SwendsenWang87}
R.~H. Swendsen and J.-S. Wang, Phys. Rev. Lett. {\bf 58}, 86 (1987).

\bibitem{Wolff89}
U.~Wolff, Phys. Rev. Lett. {\bf 62}, 361 (1989).

\bibitem{Liang92}
S.~Liang, Phys. Rev. Lett. {\bf 69}, 2145 (1992).

\bibitem{CataudellaFranzese94}
V.~Cataudella, et al., Phys. Rev. Lett. {\bf 72}, 1541 (1994).

\bibitem{MatsubaraSato97}
F.~Matsubara, et al., Phys. Rev. Lett. {\bf 78}, 3237 (1997).

\bibitem{KandelBenav90}
D.~Kandel, R.~Ben-Av and E.~Domany, Phys. Rev. Lett. {\bf 65}, 941 (1990).

\bibitem{HukushimaNemoto96}
K.~Hukushima and K.~Nemoto, J. Phys. Soc. Jpn. {\bf 65}, 1604 (1996).

\bibitem{MarinariParisi98c}
E.~Marinari, G.~Parisi, and J.J. Ruiz-Lorenzo,
in {\it Spin Glasses and Random Fields},
edited by A.~P. Young (World Scientific, Singapore, 1998).

\bibitem{SwendsenWang86}
R.~H. Swendsen and J.-S. Wang, Phys. Rev. Lett. {\bf 57}, 2607 (1986).

\bibitem{EdwardsAnderson75}
S.~F. Edwards and P.~W. Anderson, J. Phys. {\bf F 5}, 965 (1975).

\bibitem{BhattYoung88}
R.~Bhatt and A.~P. Young, Phys. Rev. {\bf B 37}, 5606 (1988).

\bibitem{LemkeCampbell96}
N.~Lemke and I.~A. Campbell, Phys. Rev. Lett. {\bf 76}, 4616 (1996).

\bibitem{ShirakuraMatsubara96}
T.~Shirakura and F.~Matsubara, J. Phys. Soc. Jpn.{\bf 65}, 3138 (1996).

\bibitem{KitataniSinada00}
H.~Kitatani and A.~Sinada, J. Phys. {\bf A 33}, 3545 (2000).

\bibitem{ShirakuraMatsubara00}
T.~Shirakura and F.~Matsubara (cond-mat/0011235).

\bibitem{SaulKardar93}
L.~Saul and M.~Kardar, Phys. Rev. {\bf E 48}, R3221 (1993).

\bibitem{RednerMachta98}
O.~Redner, J.~Machta and L.~F.~Chayes, Phys. Rev. E {\bf 58}, 2749 (1998).

\bibitem{HoudayerMartin99}
J.~Houdayer and O.~C. Martin, Phys. Rev. Lett. {\bf 83}, 1030 (1999).

\bibitem{McMillan84}
W.~L. McMillan, J. Phys. {\bf C 17}, 3179 (1984).

\bibitem{HoudayerMartin00b}
J.~Houdayer and O.~C. Martin, Europhys. Lett. {\bf 49}, 764 (2000).
\end{thebibliography}
\end{document}